\documentclass[journal]{IEEEtran}
\usepackage{cite}
\usepackage{hyperref}
\usepackage{amsmath,amssymb,amsfonts}
\usepackage{algorithmic}
\usepackage{graphicx}
\usepackage{textcomp}
\usepackage{soul}
\usepackage{xcolor}
\usepackage{braket}
\usepackage{enumitem}  
\usepackage{multirow}
\usepackage[qm]{qcircuit}
\pdfmapfile{+sansmathaccent.map}
\usepackage[linesnumbered,ruled,vlined]{algorithm2e}
\newcommand{\etal}{\emph{et al. }}
\makeatletter
\renewcommand*\env@matrix[1][c]{\hskip -\arraycolsep
  \let\@ifnextchar\new@ifnextchar
  \array{*\c@MaxMatrixCols #1}}
\makeatother

\begin{document}
\title{Towards an Optimal Hybrid Algorithm for EV Charging Stations Placement using Quantum Annealing and Genetic Algorithms}

\makeatother
\DeclareRobustCommand*{\IEEEauthorrefmark}[1]{%
\raisebox{0pt}[0pt][0pt]{\textsuperscript{\footnotesize\ensuremath{#1}}}}
  

\author{\IEEEauthorblockN{Aman Chandra\IEEEauthorrefmark{2}, Jitesh Lalwani\IEEEauthorrefmark{1,2}, and
Babita Jajodia\IEEEauthorrefmark{3}} \\
\IEEEauthorblockA{\IEEEauthorrefmark{1}Artificial Brain Tech Inc, 2055 Limestone RD, STE 200-C, Wilmington, Delaware, USA 19808 \\
\IEEEauthorrefmark{2}Artificial Brain Technology (OPC) Private Limited, Pune, India 411057\\
}
\IEEEauthorblockA{\IEEEauthorrefmark{3}Department of Electronics and Communication Engineering,\\
Indian Institute of Information Technology Guwahati, India\\
Email: \{aman.chandra@artificialbrain.in, jitesh.lalwani@artificialbrain.us, babita@iiitg.ac.in\}
}}
\maketitle

\begin{abstract}
Quantum Annealing is a heuristic for solving optimization problems that have seen a recent surge in usage owing to the success of D-Wave Systems. This paper aims to find a good heuristic for solving the Electric Vehicle Charger Placement (EVCP) problem, a problem that stands to be very important given the costs of setting up an electric vehicle (EV) charger and the expected surge in electric vehicles across the world. The same problem statement can also be generalized to the optimal placement of any entity in a grid and can be explored for further uses. 
Finally, the authors introduce a novel heuristic combining Quantum Annealing and Genetic Algorithms to solve the problem. The proposed hybrid approach entails seeding the genetic algorithms with the results of quantum annealing. Experimental results show that this method decreases the minimum distance from Points of Interest (POI) by $42.89\%$ compared to vanilla quantum annealing over the sample EVCP datasets.   
\end{abstract}
\begin{IEEEkeywords}
Electric Vehicle, Optimization, D-Wave Systems, Quantum Annealing, Genetic Algorithm, Quadratic Unconstrained Binary Optimization (QUBO)
\end{IEEEkeywords}
\section{Introduction}
Electric Vehicle Charger Placement (EVCP) is an optimization problem that is sure to become of great interest in the upcoming years, given the increase in electric vehicles, and the costs associated with setting up an electric charger. However, this problem is non-deterministic polynomial time (NP) hard, and becomes intractable with an increase in the number of charging stations to be placed.

Optimal EVCP is a problem that the authors suspect will be of great interest in the near future, owing to the fact that electric vehicles are just starting to catch up in popularity, meanwhile the infrastructure to support them is just starting to be built.
Few research works are available till date on EVCP problem \cite{FREDRIKSSON201977}, \cite{HE2018641}, and to the best of the authors' knowledge, there is no existing literature or work done yet on using the newly available strategy of quantum annealing \cite{2010}.
To explore this area further, the authors referred the work done in \cite{FREDRIKSSON201977} and \cite{HE2018641} and decided to model a basic version of the problem on quantum annealing of D-Wave computers.
The only reference available is a GitHub repository \cite{github_ev_charger} outlining a very basic and non-performant example. the authors took the same and made few developments to improve the performance greatly, and incorporated a few novel techniques in doing so. The reason the authors we believe it important to have a workable quantum solution is that annealing is a method that shows promise to greatly outperform classical heuristics when it comes to optimization problems \cite{2010}. Quantum computing hardware is still in an intermediate stage, but the authors believe it necessary to accelerate the adoption process by working towards useful solutions feasible on the currently available D-Wave hybrid solvers \cite{leap_hybrid_solver}.
\vspace{-0.17ex} \\
~~~~~~~~~~~~~~~~The motivation behind this paper was to improve on the method for solving the EVCP problem proposed by Pagany \etal \cite{su11082301}. The focus of this paper is threefold. First, to make any changes to the only quantum solution so as to improve the result. Then to find a good machine learning algorithm for solving the problem when seeded randomly, and finally to integrate the two heuristics together to see if there are any advantages in computation time, robustness and optimality of solution.
\vspace{-0.22ex}\\
~~~~~Given the limitations of current quantum hardware, an entirely quantum approach to solving a problem is not practical. However, this work stands to show that quantum computing can be utilised despite this hardware limitation to improve results of the classical heuristics that are currently being used to solve similar problems. The authors suggest that this is the best way to utilise quantum computing in the current era, and we wish to explore more use cases where this approach improves performance in future research.
\vspace{-0.22ex}\\
The paper is organised as follows: Section \ref{quantum_annealing_section} gives a brief overview of quantum annealing using D-Wave computers and evolutionary Genetic Algorithms. Section \ref{proposed_hybrid_approach} gives a detailed description of the proposed hybrid algorithm using quantum annealing and genetic algorithm towards an optimal electric vehicle charging stations placement problem. Section \ref{performance metric} discusses about the performance metric that establish the proposed algorithm along with experimental results and discussions in Section \ref{results_and_discussions}. Section \ref{conclusions_fw} concludes the paper with future works.
\vspace{-1.99ex}
\section{Background on D-Wave Quantum Annealing and Genetic Algorithms}
\label{quantum_annealing_section}
\subsection{Quantum Annealing on D-Wave Computers}
Quantum annealing is a meta-heuristic for finding the global minima of an objective function. The quantum advantage stems from the ability to explore multiple candidate solutions in parallel, as well as the ability to quantum tunnel, overcoming energy walls between two local minima, allowing for easier traversal of the energy landscape \cite{dwave_website}. 

Using the D-Wave methodology, the problem is encoded as a Quadratic Unconstrained Binary Optimization (QUBO) problem, and quantum annealing is used to find the solution to this QUBO formulation. The objective function $f(x)$ in a QUBO problem is written in the form of
\begin{equation}
\label{qubo_objective_function}
f(x) = \sum_{i=1}^{N} \sum_{j=1}^{i} q_{ij}x_{i}x_{j}
\end{equation}
where, $X$ is a vector of binary variables $x$, and $q_{ij}$ are the weights of the $N \times N$ QUBO matrix. The weights can be written in the form of either a symmetric matrix or an upper triangular matrix that we call the $N \times N$ QUBO matrix. 

The result of the quantum annealing is the value of $X$ that minimizes the objective function $f(x)$. The challenge with Quantum annealing lies in finding the correct values of QUBO matrix to efficiently represent the problem that the authors are aiming to solve. 
\subsection{Genetic Algorithms} 
Genetic algorithms are a well known heuristic for solving optimization problems that are difficult to solve precisely \cite{ELSAYED201457}. The general outline of a genetic algorithm can be described as follows:
\begin{enumerate}
    \item Define an individual that carries the necessary information in it's genes to be a solution to the optimization problem
    \item Create a population of these individuals
    \item Define a method for breeding these individuals
    \item Define a method for mutation in the population
    \item Define a fitness function that describes how good the solution represented by the genes are
    \item Carry out an evolutionary simulation that mimics the concept of natural selection, where the population is updated every generation following the mutation and breeding methods mentioned above.
\end{enumerate}
Then, the population of the final generation is used as a solution to the problem.
\section{Proposed Hybrid Algorithm for EVCP using Quantum Annealing and Genetic Algorithms}
\label{proposed_hybrid_approach}
\subsection{The EVCP Problem}
\begin{figure}[!t]
\centering
\includegraphics[height=5cm,width=7cm]{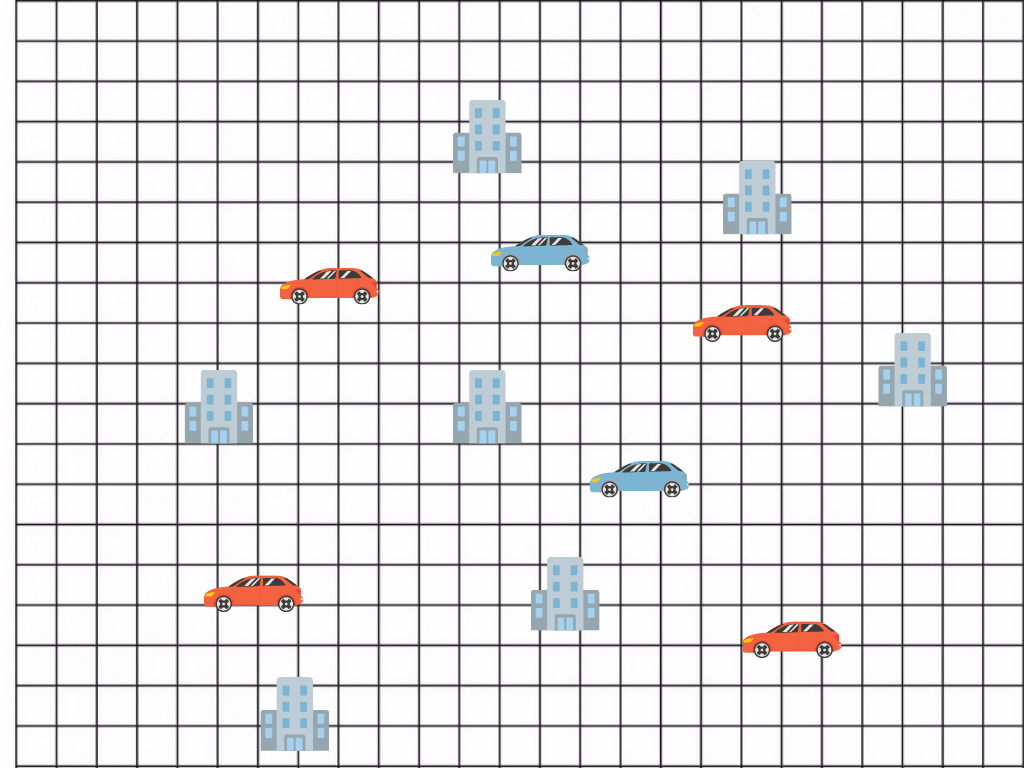}
\caption{Picture showing an example of EVCP, where buildings are Points of Interest (POI), blue cars are old chargers and red cars are new chargers}
\label{figure_3}
\end{figure}

The EVCP can be defined by setting up a grid on which the Points of Interest (POI) and already existing electric chargers are set up on the nodes. The proposed algorithm then places down new charging points in an optimal manner. For the sake of this research work, optimality has been defined as minimizing travel distance to a charger for all points of interest on the grid. As shown in Fig. \ref{figure_3}, the algorithm will decide where to place the red cars (new charging points), to optimize the scoring metric that will be explained in Section \ref{scoring_metric}. 
\subsection{Proposed Quantum Annealing Approach}
This proposed quantum annealing approach is inspired by the QUBO formulation presented in \cite{su11082301} that aims to optimize three objectives:
\begin{enumerate}
    \item Minimize distance from Points of Interest (POI)
    \item Maximize distance from previous existing charger locations
    \item Maximize distance between the new charger locations
\end{enumerate}
Based on these objectives, the three constraints are chosen to try to place the chargers close to the POI, without placing them trivially close to each other or previously present chargers. The reason for including all three is to distribute the points to maximize coverage of POI for the given number of chargers to be placed. It must be noted that in the QUBO formulation, the centroid of all POI are considered for minimization, and similarly for previously existing chargers as well as new chargers. 

Mathematically, the three constraints ($H_{1}$, $H_2$ and $H_3$) referring to the three objective functions can be represented as follows:
\begin{equation}
\label{first_constraint}
H_{1} = +\sum_{i=1}^{N} x_{i}d_{i}^{p} 
\end{equation}
\begin{equation}
\label{second_constraint}
H_{2} = -\sum_{i=1}^{N} x_{i}d_{i}^{c} 
\end{equation}
\begin{equation}
\label{third_constraint}
H_{3} = -\sum_{i=1}^{N} x_{i}d_{i}^{l} 
\end{equation}
with the values of $d_{i}^{p}$, $d_{i}^{c}$ and $d_{i}^{l}$ given as
\begin{equation}
\label{d_i_p_equation}
    d_{i}^{p} = \sum_{k=1}^{n_{POI}} dist(POI_{k} - x_{i})^{2} 
\end{equation}
\begin{equation}
    d_{i}^{c} = \sum_{k=1}^{n_{charger}} dist(charger_{k} - x_{i})^{2}
\end{equation}
\begin{equation}
    d_{i}^{l} = \sum_{k=1}^{n_{locations}} dist(x_{k} - x_{i})^{2} 
\end{equation}
Here, $N$ is the no. of points, $POI$ are the Points of Interest, $charger$ refers to the chargers that are already present, thus $n_{charger}$ is the number of chargers, $n_{locations}$ refers to the number of available locations to place a new charger, $m$ is the no. of new chargers we want to place and $dist$ refers to euclidean distance between the two points.

The last constraint where the number of chargers must be equal to a value $m$ can be given as
\begin{equation}
     H_{4} = (\sum_{i=1}^{N}x_{i}) - m 
\end{equation}

The final QUBO formulation can be given by
\begin{equation}
\label{final_qubo_equation}
    H_{final}= \sum_{i=1}^{N}\lambda_iH_i
\end{equation}
Please note that the parameters $\lambda$ constitute the trade-off between different optimization functions. A careful balance of all the constraints ($H_{1}$, $H_2$, $H_3$ and $H_4$) is required to achieve the aim of the scoring metric described in Section \ref{scoring_metric}. 
\subsection{Inclusion of Entropy}
An error observed with the initial QUBO formulation is that it rewarded extreme minimization to a few POI at the expense of other POI. This has been combated by changing the distance minimization in $H_{1}$ to a distance entropy minimization. The motivation for this idea stems from information entropy given by
\begin{equation}
\label{entropy_equation}
    H(X) = \sum_{}{} P(x_{i})\log P(x_{i}) 
\end{equation}
where $X$ is a discrete distribution, $H(X)$ is the entropy of the distribution and $P(X)$ is the probability distribution of the variable $X$.

For the proposed hybrid algorithm, we replaced \eqref{d_i_p_equation} with $H(d^{p})$, by taking our distribution to be a $softmax$ distribution, and used this new equation to calculate \eqref{first_constraint}. The chosen probability distribution can be thought of as a $softmax$ over the distances of the POI from each possible charger location, and minimizing the value of this entropy instead of just the distances from the POI has led to significantly improved results as compared to before this change on the datasets. For the rest of this paper, Quantum Annealing has been used with this entropic version of \eqref{d_i_p_equation} and all results have been calculated using this, as they were too poor to be considered otherwise.
\begin{figure}[!t]
\centering
\includegraphics[height=5.5cm,width=8.cm]{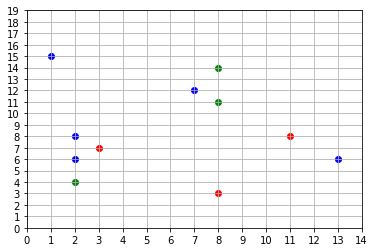}
\caption{Sample Output using only quantum annealing. The blue, red and green colours in the graph indicates POI, old charging stations and new charging stations respectively, with the score of 64.0}
\label{figure_exp_1}
\end{figure}
\begin{figure}[!t]
\centering
\includegraphics[height=5.5cm,width=8.5cm]{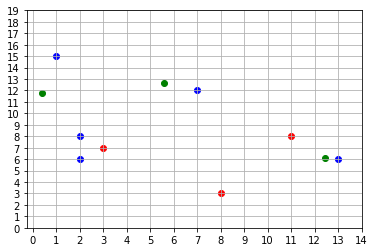}
\caption{Sample Output where a GA was seeded with quantum annealing. The colours- blue, red and green indicates POI, old charging stations and new charging stations respectively with the score of 17.74}
\label{figure_exp_2}
\end{figure}
\subsection{Integration of Evolutionary Genetic Algorithm with Quantum Annealing}
The output from the Quantum annealing, although significantly improved after introducing an entropic loss function, and optimization of hyper-parameters, still has much room for improvements. 
The authors introduce a method of using evolutionary genetic algorithms to optimize the output received from the quantum annealing. Research findings show that seeding a genetic algorithm with the results of quantum annealing improves on the result of either on its own to solve the problem.

The authors first explored the uses of genetic algorithms in solving the problem of EVCP directly, and found out that the use of genetic algorithms brings two advantages as compared to quantum annealing as follows:
\begin{enumerate}
    \item Reduced time to solution - it completes a basic search much quicker
    \item Easy transformation into a continuous problem - The quantum annealing solves this as a discrete problem, using evolutionary algorithms however, it is possible to turn the same into a continuous problem.
\end{enumerate}
However, the problem becomes very chaotic to solve using a randomly seeded genetic algorithm when the number of new chargers increases. The authors theorize that this is due to the energy landscape being scattered with local minima, especially if the search space is assumed to be continuous. Since the heuristic used involves a random mutation and crossover approach, it is quite difficult for it to traverse this landscape. The proposed solution is to use a quantum annealing to search the space discretely and traverse over the local minima in the energy landscape of the problem. Then, the genetic algorithm can take outputs of the quantum annealing as a seed and optimize it further. This optimization is doubly useful as it turns the problem discrete, which allows better solutions than possible in a discrete space. This also allows the genetic algorithm to search the local area highlighted by the annealing to improve on it if possible.

Fig. \ref{figure_exp_1} shows the output from just quantum annealing, while Fig. \ref{figure_exp_2} shows the output from the proposed quantum annealing + genetic algorithm. These graphs are plotted using the sample EVCP~(5,3,3) dataset. As it can be seen from Fig. \ref{figure_exp_2}, the score is lower with respect to Fig. \ref{figure_exp_1}. The lower the score is, the better the placement of charger is. The placement of the chargers also looks to be better intuitively. 

This strategy does have some drawbacks that the authors would like to point out. The algorithm will fail entirely if the genetic algorithm used is not capable of solving the problem independently, due to either a poor crossover and mutation strategy, or a bad choice of hyper-parameters. It might be advisable to perform some kind of search to find the best parameters possible, and extensive experimentation needs to be done to find the right crossover and mutation strategies, depending on the specific problem. 
\section{Performance Metric}
\label{performance metric}
\subsection{Scoring Metric}
\label{scoring_metric}
The proposed hybrid algorithm aims to solve the charger placement problem by minimizing the sum of minimum distance to a charger from each POI by $\sum min(D_{k}^{i})$, where $D_{k}^{i}$ is the array of distances of all $k$ charger locations from the $i^{th}$ point of interest. This hopes to place new chargers such that they are not redundant due to other chargers existing close by, but are still close to POI, reducing travel time and cost for reaching a charger. This scoring metric however, did not account for robustness and thus the score used finally was the sum of the mean and variance of the above the proposed scoring metric over the no. of runs.
\subsection{Disparity of QUBO formulation with Scoring Metric}
\label{disparity_scoring_metric}
It is reasonable to question why the authors did not create a QUBO that directly minimized what the scoring metric described above measures. The problem was the involvement of a $min()$ function in the scoring metric. This cannot be easily translated into a binary quadratic problem, and the increased computation during the QUBO generation due to the added computational complexity would make the process extremely inefficient. Therefore, the objective functions were chosen such that $(i)$ they were simple and efficient to code in and $(ii)$ could be used to give a result that minimized our scoring metric when the right lambda parameters were chosen. Hence, there was no testing done using that formulation, however if an efficient method to encode the scoring metric into a QUBO can be found, the authors suspect the results would be better than what we have achieved.
\subsection{Parameter Search}
The Lagrange parameters for the QUBO were optimized using a Bayesian search which aimed to minimize the score form the scoring metric described in Section \ref{scoring_metric} and \ref{disparity_scoring_metric}. This parameter search is a necessary step since the performance of the algorithm is very dependent on these parameters. However, this is a very computationally expensive step, and thus it is not scalable to carry out this search every time the algorithm is run. Optimal parameters change based on the problem size and number of points quite drastically, hence the same values can not be used except for specifically chosen problem sets. 
\begin{table}[!t]
    \centering
     \label{comparison_table}
     \caption{Comparison Table of the Proposed Hybrid Algorithm integrating Quantum Annealing and Genetic Algorithm over existing only QA and GA on our sample EVCP datasets}
     \begin{tabular}{|c|p{1.5cm}|p{1.5cm}|p{1.5cm}|}
    \hline
         \multirow{2}{*}{\textbf{Sample EVCP Dataset}} &  \multicolumn{3}{c|}{\textbf{Score}} \\
         \cline{2-4}
          & Only QA$^\dagger$ & Only GA$^\ddagger$ & \textbf{ Proposed$^\ddagger$$^\ddagger$} \\
         \hline
         EVCP~(5,2,3) & 40 & 91.40 & \textbf{23.70} \\
         \hline
         EVCP~(5,3,3) & 63 & 43.77 & \textbf{34.80} \\
         \hline
         EVCP~(6,3,3) & 105 & 105 & \textbf{48.81} \\
         \hline
         EVCP~(9,3,3) & 172 & 185.43 & \textbf{60.44} \\
         \hline
         EVCP~(10,3,3) & 400 & 945.02 & \textbf{339.16} \\
         \hline
         EVCP~(15,1,3) & 781 & 1267.95 & \textbf{467.37} \\
         \hline 
         EVCP~(20,4,3) & 1071.66 & 1872.87 & \textbf{624.77} \\
         \hline 
         EVCP~(20,4,4) & 5014 & 6488.56 & \textbf{2922.91} \\
         \hline 
         EVCP~(20,3,3) & 1402 & 1868.76 & \textbf{1115.66} \\
         \hline
         \multicolumn{4}{c}{$^\dagger$QA: Quantum Annealing, $^\ddagger$GA: Genetic Algorithm} \\ 
         \multicolumn{4}{c}{$^\ddagger$$^\ddagger$Proposed: Integration of QA and GA} 
    \end{tabular}
    \label{results_table}
\end{table}
\section{Experimental Results and Discussions}
\label{results_and_discussions}
\begin{figure}[!t]
\centering
\includegraphics[height=5.5cm,width=8.5cm]{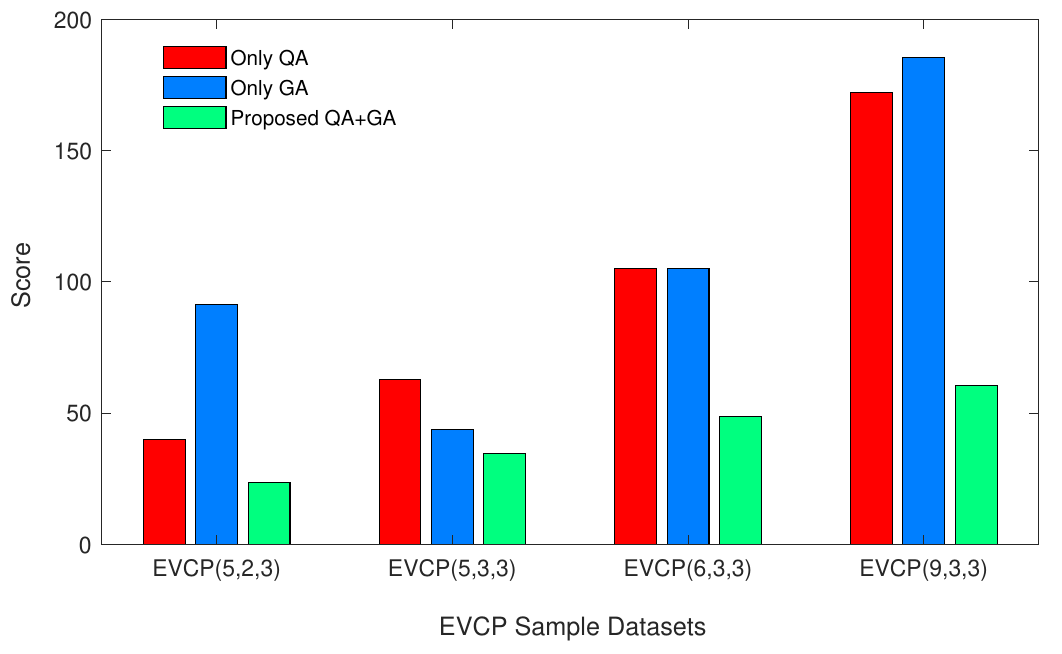}
\caption{Comparison Results of Only QA, Only GA and Proposed QA+GA for EVCP sample datasets: EVCP~(5,2,3), EVCP~(5,3,3), EVCP~(6,3,3), EVCP~(9,3,3). The lower the score, the better the performance}
\label{figure_exp_3}
\end{figure}

\begin{figure}[!t]
\centering
\includegraphics[height=5.5cm,width=8.5cm]{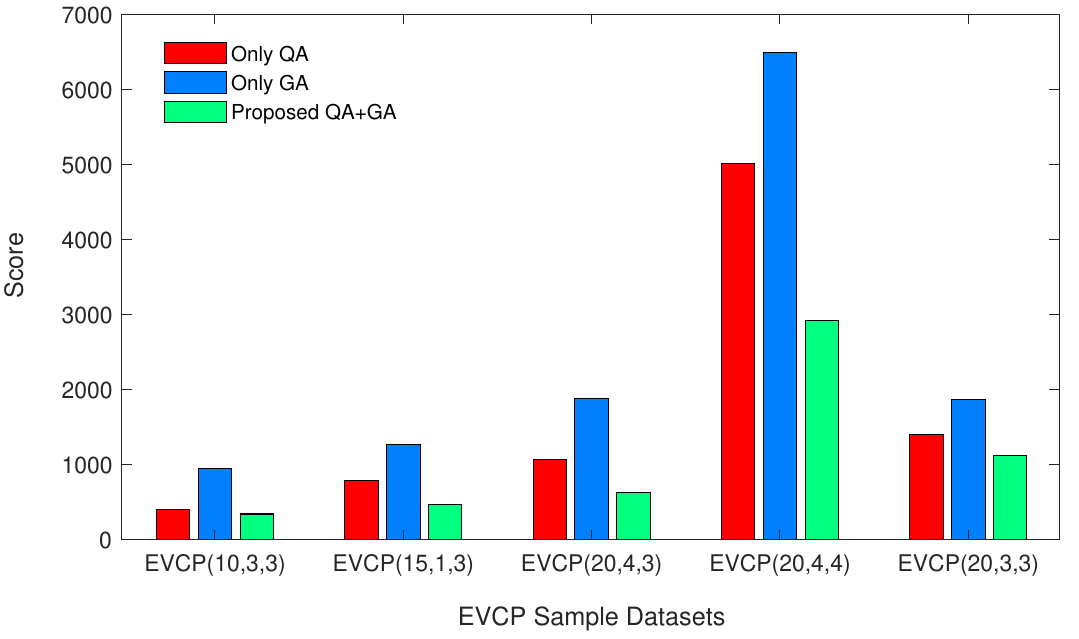}
\caption{Comparison Results of Only QA, Only GA and Proposed QA+GA for EVCP sample datasets: EVCP~(10,3,3), EVCP~(15,1,3), EVCP~(20,4,3), EVCP~(20,3,3), EVCP~(20,4,4). The lower the score, the better the performance}
\label{figure_exp_4}
\end{figure}

Experimental evaluations were performed using D-Wave Quantum Annealing along with the two other techniques as follows:
\begin{enumerate}
    \item Only Quantum Annealing
    \item Only Genetic Algorithm
    \item Integration of Quantum Annealing and Genetic Algorithm
\end{enumerate}
The first evaluation is based on the score described in Section \ref{scoring_metric} of the aforementioned techniques over a few differently sized datasets. 
The authors have named the sample datasets as $EVCP~(n_{POI}, n_{ch\_old}, n_{ch\_new})$, where $n_{POI}$, $n_{ch\_old}$ and $n_{ch\_new}$ are the no. of POI, no. of existing charging stations and 
no. of new charging stations to be placed respectively. Experiments were performed over a few sample data sets with grid size of 15$\times$20, 30$\times$30 and 100$\times$100 are as follows:
\begin{enumerate}
\item 15$\times$20 grid: EVCP~(5,2,3), EVCP~(5,3,3), EVCP~(6,3,3), EVCP(9,3,3)
\item 30$\times$30 grid: EVCP~(10,3,3),EVCP~(15,1,3),EVCP~(20,4,3), EVCP~(20,3,3)
\item 100$\times$100 grid: EVCP~(20,4,4) 
\end{enumerate}
Table \ref{results_table} illustrates the score of the proposed hybrid algorithm (integration of quantum annealing and genetic algorithm) over only quantum annealing and only genetic algorithm on the sample EVCP datasets. It is also clear from Fig. \ref{figure_exp_3} and Fig. \ref{figure_exp_4} that the advantage provided by the genetic approach is greater when the grid size is larger. The authors suspect this isn't due to the GA performing better on larger datasets, but rather due to the limited size of the Bayesian search in finding the Lagrange parameters for the quantum annealing. This is mostly a temporal constraint, as the run time for the search was very high for the 100$\times$100 grid EVCP~(20,4,4) sample dataset, and it might have even been possible to find an exact solution in that time. With our proposed quantum annealing + genetic algorithm approach, however, it is possible to get good results without an extensive parameter search, and that saves on a lot of time. It should also be mentioned that the search for the right parameters was run on QBSolv \cite{qbsolv}, a simulator provided by D-Wave Systems. This is due to the limited compute time available on the Leap hybrid solvers \cite{leap_hybrid_solver}.

\begin{figure}[!t]
\centering
\includegraphics[height=5.0cm,width=8.5cm]{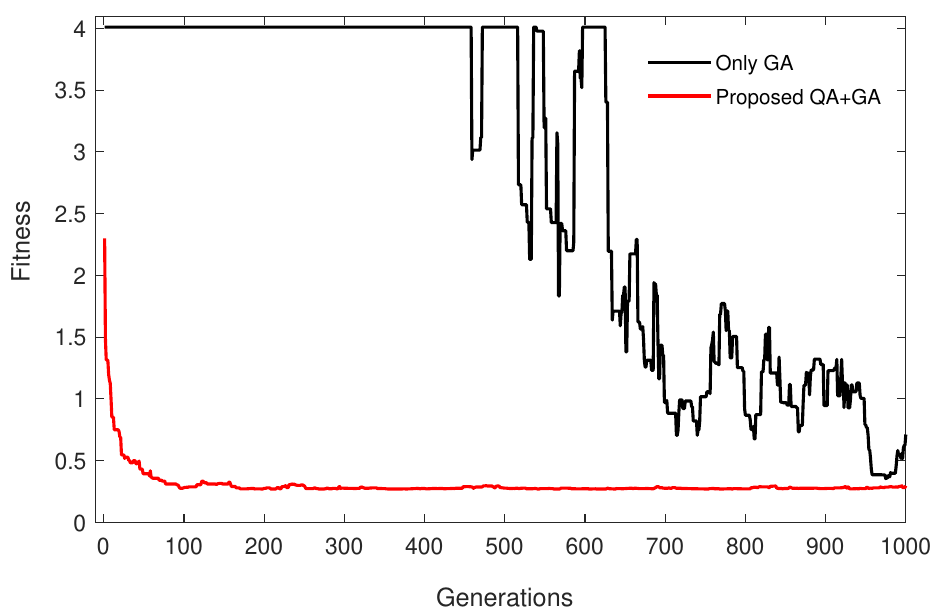}
\caption{Plot of fitness vs generation using (a) Only Genetic Algorithm and (b) Integration of Quantum Annealing and Genetic Algorithm}
\label{figure_exp_5}
\end{figure}
After confirming the increase in performance (Fig. \ref{figure_exp_3} and Fig. \ref{figure_exp_4}), the best score per epoch were also calculated and plotted for the randomly seeded and Quantum annealing seeded algorithms as shown in Fig. \ref{figure_exp_3} and Fig. \ref{figure_exp_4}. The largest dataset taken by us for the experimental evaluations contains a $100 \times 100$ grid and 20 POI.

It can be clearly illustrated from Fig. \ref{figure_exp_5} that the training process is much more stable and the convergence is much quicker for the Quantum annealing seeded process. The disadvantage of this proposed hybrid method however is that the total time is the sum of both quantum annealing and the genetic algorithm, so in situations which are extremely time sensitive, this approach is not the best. However, in larger data sets, the time taken for the genetic algorithm is trivial compared to the quantum annealing part of the hybrid algorithm.

It should be noted that the genetic algorithm was run for 100 generations when seeded by the quantum annealing, while for 1000 generations when randomly seeded for these results (Table \ref{results_table}). All other parameters were kept constant.

The authors tested the performance of these proposed strategies on the EVCP sample datasets, and saw that the scores were significantly improved as compared to a vanilla quantum annealing, with an average improvement in score of $42.89\%$ across all the data sets tested on, based on minimum distance to POI. For each dataset, the score for each computing method was the average over five runs. This is quite a great result when looked at face value, however it has to be noted that the Bayesian search could not be performed extensively on larger datasets, which led to a decrease in performance of Quantum Annealing. This is because it is not possible to run the Bayesian search effectively for larger grid sizes, due to extremely high computation times. The other factor to note here is the lack of research done in Quantum annealing with regards to the EVCP, which points to the fact that improvements on the purely quantum end are sure to be made there.
\section{Conclusion and Future Works}
\label{conclusions_fw} 
This work started with the aim of improving the performance of quantum annealing in solving the EVCP problem. In trying to do so, the authors have come up with strategies that might be usable beyond just this problem and can be applicable to multiple NP hard problems. Based on experimental results, the authors mentioned findings that seeding a evolutionary genetic algorithm with the results of a quantum annealing provides a decrease in minimum distance from POI of $42.89\%$ compared with QA and $57.54\%$ compared to a randomly seed GA. 
The algorithm outlined above shows great promise in terms of making quantum annealing usable in the near term. The authors therefore think that both the inclusion of an entropic optimization function in the QUBO formulation, as well as the seeding of genetic algorithm using quantum annealing needs to be tested on other problems that can be solved using quantum annealing. 
\section{Acknowledgements}
The authors would like to thank D-Wave Systems for continuous support and providing our team the platform to perform experiments using D-Wave Quantum Computers. Also a big thanks to Naman Jain from Artificial Brain Team for helping review the paper and bringing up great points during the process. 
\bibliographystyle{IEEEtran}
\bibliography{references}
\end{document}